# Plasmons in MoS2 studied via Experimental and Theoretical Correlation of Energy Loss Spectra

Eoin Moynihan[1]

Stefan Rost[2]

Eoghan O'Connell[1]

Quentin Ramasse[4]

Christoph Friedrich[3]

Ursel Bangert[1]

1: TEMUL, Department of Physics, School of Sciences & Bernal Institute, University of Limerick, Limerick, Ireland.

2: Peter Grünberg Institut and Institute for Advanced Simulation Forschungszentrum Jülich and JARA, 52425 Jülich, Germany and RWTH Aachen University, 52062 Aachen, Germany

3: Peter Grünberg Institut and Institute for Advanced Simulation Forschungszentrum Jülich and JARA, 52425 Jülich, Germany

4: SuperSTEM Laboratory, SciTech Daresbury Campus, Daresbury WA4 4AD, United Kingdom

This paper takes a fundamental view of the electron energy loss spectra of monolayer and few layer $MoS_2$. The dielectric function of monolayer $MoS_2$ is compared to the experimental spectra to give clear criteria for the nature of different signals. Kramers-Krönig analysis allows a direct extraction of the dielectric function from the experimental data. However this analysis is sensitive to slight changes in the normalisation step of the data pre-treatment. Density functional theory provides simulations of the dielectric function for comparison and validation of experimental findings. Simulated and experimental spectra are compared to isolate the $\pi$ and $\pi + \sigma$ surface plasmon modes in monolayer $MoS_2$. Single-particle excitations obscure the plasmons in the monolayer spectrum and momentum resolved measurements give indication of indirect interband transitions that are excited due to the large convergence and collection angles used in the experiment.

## Introduction

Layered anisotropic materials, i.e. graphite, MoS₂, h-BN, have been studied extensively for their dielectric and plasmonic properties over several decades [1–5]. Nanotubes made from these materials pushed further research into plasmonic properties [6,7]. After the isolation of single sheets (monolayers) of graphene, two-dimensional (2D) materials have been exfoliated from their bulk counterparts and launched new interest in their plasmonic properties [8].

Plasmons are the collective oscillation of valence or conduction electrons in a material. Classically, this is described by the complex dielectric function of a given material where the real component describes the transmission of electromagnetic waves through the medium and the imaginary component describes single-particle excitations (i.e interband transitions) [9]. The criterium for collective excitations such as plasmons is that the real part of the dielectric function crosses 0 with a positive slope. In the case where plasmons are excited at a dielectric/metal interface, where the dielectric function of the dielectric is $\varepsilon_1 > 0$ and the dielectric function of the metal is $\varepsilon_2 < 0$, a surface plasmon follows the criteria [9,10]:

$$\varepsilon_1(\omega) + \varepsilon_2(\omega) = 0 \tag{1}$$

And for a bulk plasmon energy of $E_P$, when the dielectric is vacuum with $\varepsilon_1 = 1$, the surface plasmon energy $E_s$ is:

$$E_s = \frac{E_p}{\sqrt{1+\varepsilon_1}} = \frac{E_p}{\sqrt{2}} \tag{2}$$

Plasmons have become the centre of intense research focus due to the many applications that can be derived from their interactions with light. Electron energy loss spectroscopy (EELS) is a commonly used technique for the investigation of plasmons at the nanoscale due to the high spatial resolution in transmission electron microscopy (TEM) and the responsiveness of plasmons to excitation via fast moving electrons. In previous decades, layered materials, such as graphite and MoS₂ were investigated for their anisotropic behaviour. The resonant collective transitions from the $\pi$ and $\sigma$ bonding states to the $\pi^*$ and $\sigma^*$ antibonding states manifest themselves as the lower energy $\pi$ plasmon and the higher energy $\pi + \sigma$ plasmon. These plasmons are referred to as interband plasmons. Some 2D materials also show plasmons arising from the collective transition of intra-band transitions. These are correspondingly called intraband plasmons and occur at much lower energies [11].

With the discovery of isolated monolayers of graphene there was a revival of interest in the plasmonic properties of these newly discovered 2D monolayers. Eberlein *et al.* investigated free-standing graphene monolayers using monochromated STEM EELS and they saw a shifting of the plasmonic peaks between few-layer and multi-layered graphene [12]. The $\pi$ and $\pi + \sigma$ plasmons in graphene seemed to shift to lower energies due to the disappearance of the bulk plasmon mode in mono- and few layer graphene leaving the surface modes as the prominent signals in the EELS spectra. In later EELS experiments, a novel plasmon mode was observed at ~3eV in doped graphene [13]. There has been a large amount of theoretical and experimental work looking into the different plasmon modes that exist for graphene with different dopant configurations [14,15].

Transition metal dichalcogenides (TMDCs) cover a wide class of materials with some being semi-metals ($HfTe_2$, $PtSe_2$, $TiSe_2$, etc.) similar to graphene and others being semiconductors ($MoS_2$, $WS_2$, etc.). Metallic TMDCs have been shown to present plasmons in the visible and infrared regions, making them attractive for optoelectronics. The low energy excitation at ~2.3 eV in thin films of $TiSe_2$ was investigated using momentum resolved EELS, and proposed to be a plexciton [16]. Plexcitons are coupled plasmon-exciton polaritons and have been observed in a number of different systems [17–19]. Other metallic TMDs such as $PtTe_2$ show lower energy Dirac intraband plasmons (~0.5 eV) [20,21]. Also, $TaS_2$, $TaSe_2$, and $NbSe_2$ have excitations known as charge-carrier plasmons (~1 eV) [22–24].

Semiconducting TMDCs have generally been closely studied due to their excitonic properties and their transitions from indirect in few layer to direct bandgaps monolayer. [25–27] Metal nanostructures can be decorated onto a TMDC to couple the plasmons in the metal to the excitons in the TMDC [28]. The plasmons intrinsic to these semiconducting TMDCs however have not received nearly as much attention. There have been some DFT studies of monolayer sheets of $MoS_2$, which is a typical representative of these TMDCs [29]. Because of the low carrier concentration of $MoS_2$ compared to the metallic 2D materials, most studies focus on doped $MoS_2$ for plasmonic properties [30]. Other approaches have focused on metallic edge states in nanostructured $MoS_2$ [31,32]. Nerl *et. al.* use a combination of experimental EELS measurements and theoretical time-dependant density functional theory (TDDFT) to observe plasmons and excitons in pristine few layer $MoS_2$ [33]. Their study shows how the plasmons and excitons change as a function of layer thickness, momentum, and distance from the edge of flakes. There are however no experimental EELS data for monolayer $MoS_2$. Computationally heavy Bethe-Salpeter Equation (BSE) was used to take into account the electron-hole interactions required to properly simulate excitons in the monolayer.

EELS experiments of pristine monolayers can be compared to simulations of an infinite monolayer sheet, ignoring edge effects. Whereas investigations of plasmons in $MoS_2$ are more focused on edge states and plasmon modes that occur due to doping, this paper intends to show the intrinsic properties of pristine monolayers of $MoS_2$.

## Materials & Methods

Pristine monolayer MoS2 samples were prepared using mechanical exfoliation from bulk single crystals, as described in previous work [34].

High angle annular dark field (HAADF) images and EELS spectra were acquired using a dedicated scanning transmission electron microscope (STEM), the aberration-corrected Nion UltraSTEM 100MC HERMES at the SuperSTEM in Daresbury. The microscope was operated at 60 kV with convergence and collection angles of 31 mrad and 44 mrad, respectively. EELS spectra were denoised using Hyperspy's principal component analysis (PCA) decomposition tools [35]. Kramers-Krönig Analysis was conducted on spectra of monolayer $MoS_2$ in Hyperspy using the function kramer_kronig_analysis with one iteration and using different thickness values for the normalisation. The zero-loss peak and plural scattering were removed before hand using the logarithmic fit between 1 – 1.5 eV and Fourier-log deconvolution tools within Digital micrograph 2.3.

Density Functional Theory (DFT) simulations were performed using FLEUR [36] and SPEX [37] with lattice constants of a=3.15 Å and c=12.3 Å for the hexagonal lattice structure. An internal structure parameter of z=0.124 was used. The dielectric function as a basis to compute the EELS spectra was calculated using the random-phase approximation with zero momentum transfer, i.e. q=0.

## Results and Discussion

Figure 1(a) shows the HAADF image of a section of MoS2 with a series of terraces of varying thickness. The thickness decreases going from the top to the bottom of the image as can be seen by the decreasing intensity in the HAADF image. Spectra were integrated over 30 pixels in 5 regions, on different terraces, marked in the HAADF image Figure 1(a) by boxes A – E. The spectra acquired from each of these terraces are plotted together in (b) and (c). The $\pi$ bulk plasmon can be seen at 8.6 eV in (b) and the $\pi + \sigma$ bulk plasmon mode is seen at around 23 eV in (c). There is a clear decrease in the intensity of the bulk modes with decreasing thickness, both in terms of absolute intensity and relative to the intensity of the excitons seen in (b) at ~1.9 eV(A exciton), ~2.1eV(B exciton), and ~3 - 3.5 eV(C exciton). The exciton values observed in these spectra match the known values from literature [26,33]. The A and B excitons can be understood as transitions at the K and K' points of the Brillouin zone, which are split in energy due to the spin orbit coupling caused by the lack of inversion symmetry [38]. The C exciton was initially reported by Mertens *et al.* and is strongly absorbing due to band nesting [38,39]. Band nesting is where energy bands in the valence band minimum(VBM) and conduction band maximum(CBM) are parallel to each other, for $MoS_2$ this occurs between the K and Γ points in the Brillouin zone. The bands being parallel gives them a broader absorption peak compared to the A and B excitons which come from parabola in the CVM and VBM at the K point. Direct interband transitions can also be seen be seen at around 5 eV, marked as D in Figure 1 (b). The peak shifts with decreasing thickness, which is likely to be due to the change in the tails of the $\pi$ bulk plasmon peak. These direct transitions originate from van Hove singularities in the band structure [40].

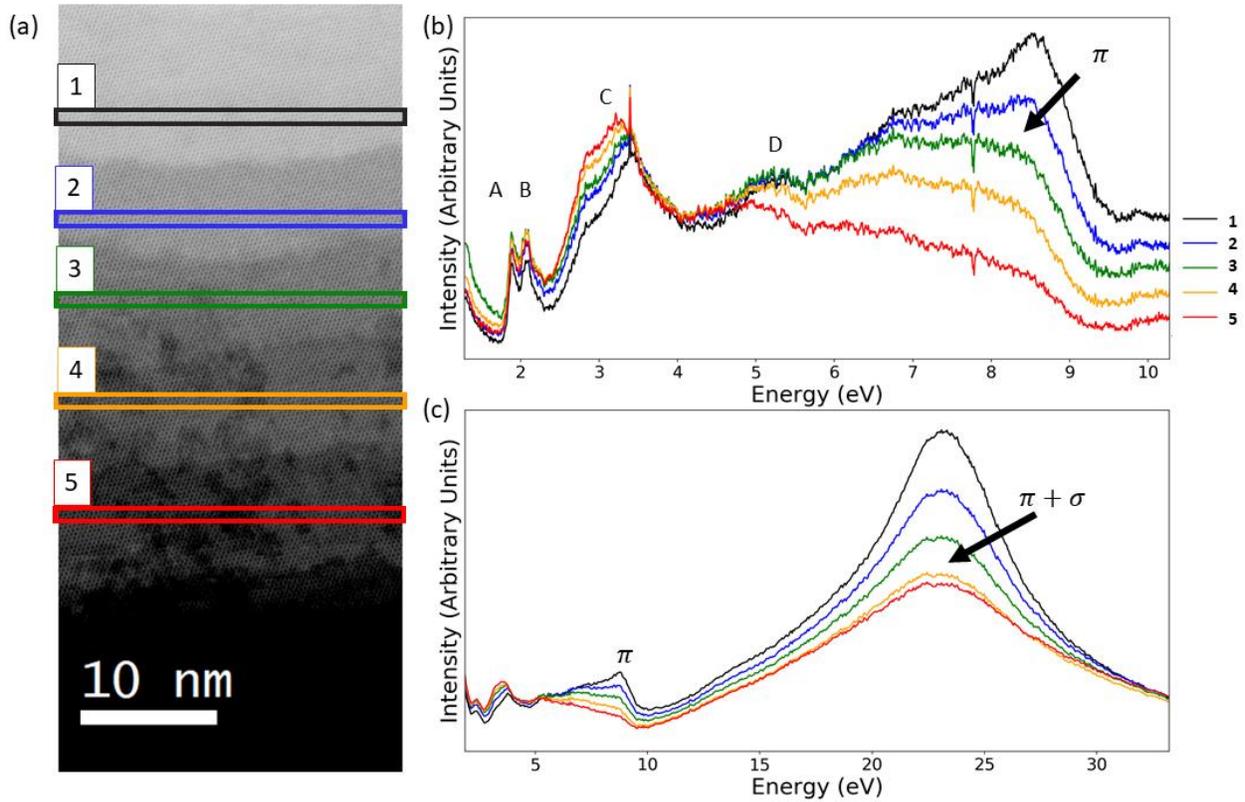

*Figure 1:* (a) HAADF image of terraces with different thicknesses in MoS2 marked at points 1-5. The coloured horizontal lines show the integration windows for the spectra extracted from the related spectrum images. (b) Spectra acquired with a dispersion of 5 meV, showing the decrease of the π plasmon (∼ 8.3 eV) as a function of decreasing thickness. (c) Spectra acquired with a dispersion of 50 meV showing the π + σ bulk plasmon (∼ 23 eV) decreasing as a function of decreasing thickness, although a large intensity is also contributed from hydrocarbon contamination.

The Begrenzungs effect is a well-known phenomenon that describes the reduction of bulk excitations in favour of surface excitations with proximity to a surface/interface [9,10]. Red-shifting of the bulk plasmon modes as the layer number decreases is clearly demonstrated in the literature and is generally attributed to coupling between plasmons in different layers. There is also a contribution to the intensity in the $\pi + \sigma$ plasmon energy range coming from hydrocarbon contamination. This contamination is attracted to the electron beam during consecutive scans and does affect the results of Figure 1 (c). However, similar results were shown in literature [33] and contamination was avoided during all other analysis by selecting clean areas of the sample from the HAADF images. To the left of both the $\pi$ and $\pi + \sigma$ bulk plasmons there are shoulders that can possibly be attributed to surface plasmon modes. The expected energy loss, from equation 2, for the $\pi$ (~5.8 eV) and $\pi + \sigma$ (~16 eV) surface plasmon match with these peaks. It would be premature to immediately refer to these peaks as surface losses. Some of the intensity in this energy range can originate from single-particle excitations, a further discussion of which is presented later in regards to monolayer $MoS_2$ using the imaginary dielectric function.

Figure 2 shows a monolayer/bilayer interface of MoS2 in a different area of the same sample. Figure 2 (b) was recorded at a higher dispersion (5 meV), which allowed for better resolution of the individual peaks in this energy range in the spectrum. Some intensity from the $\pi$ bulk mode at ~8.3 eV isstill present in the bilayer, while this mode has practically disappeared in the monolayer. The peaks in the low loss spectrum between ~4-8 eV seem to be a convolution of the interband transitions and the $\pi$ surface plasmon mode. Figure 2(c) shows a lower dispersion (50 meV) view of the low loss spectra in order to show the $\pi + \sigma$ plasmon. The bulk and surface modes of the $\pi + \sigma$ plasmon are more intense in the bilayer.

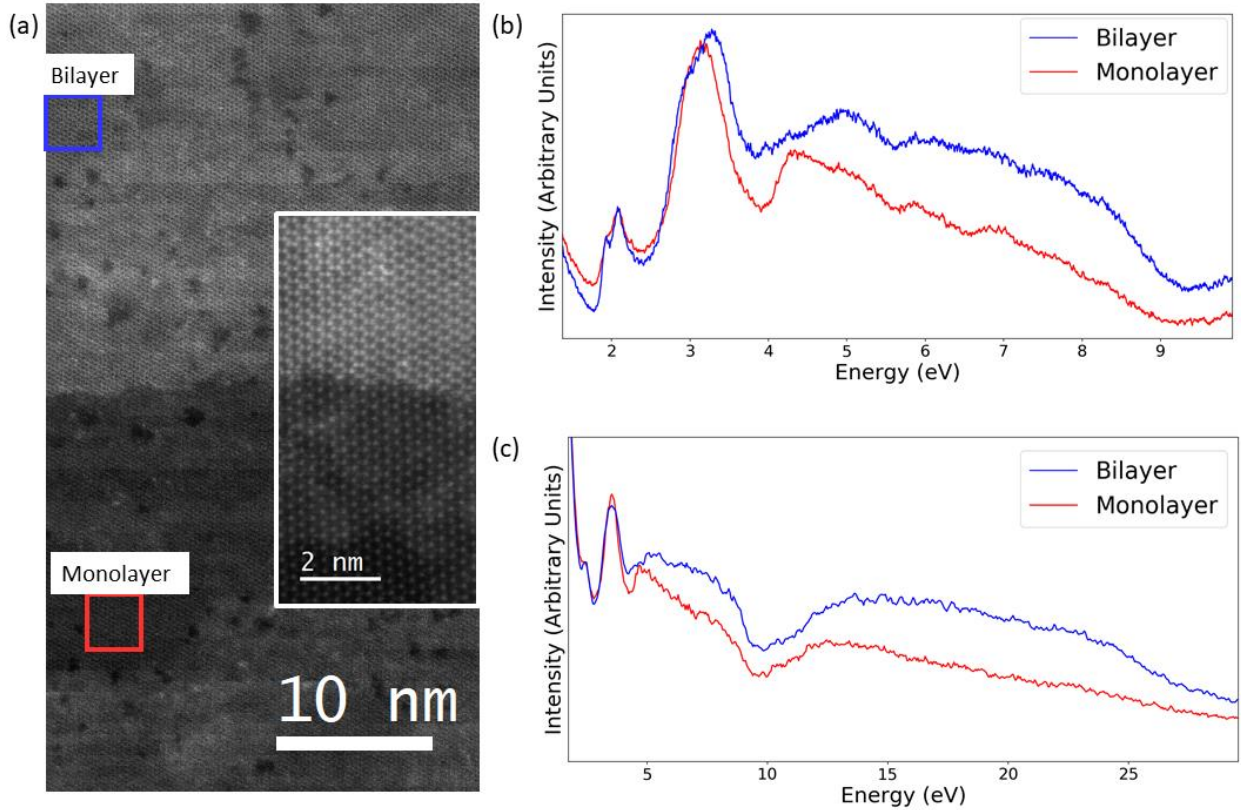

*Figure 2:* (a) HAADF image of monolayer/bilayer interface in MoS2. Spectra were integrated in regions 1 and 2 where there is no contamination (white patches) or electron beam damage (dark holes) that accumulated during multiple scans of the area. Inset shows contrast between Mo and S columns in monolayer and bilayer sections. The (b) $\pi$ bulk plasmon ($\approx 8.3$) and (c) $\pi + \sigma$ bulk plasmon ($\approx 23$) disappear for monolayer $MoS_2$

In order to draw meaningful conclusions from the experimental energy loss spectra they can be compared to the dielectric function. The dielectric function is a complex function where the real part describes the propagation of waves through a medium, while the imaginary part describes the absorption due to single-particle excitations. The dielectric function can be determined experimentally from the EELS spectrum using Kramers-Krönig analysis or theoretically via DFT simulations.

Kramers-Krönig analysis (KKA) is colloquially used as a term for the procedure used to calculate the dielectric function from an experimental EELS function. More specifically, the Kramers-Krönig transformation only refers to the last part of the procedure where the real part of the inverted dielectric function is derived from the loss function using the following equation:

$$Re\left[\frac{1}{\varepsilon(E)}\right] = 1 - \frac{2}{\pi}P\int_0^\infty Im\left[\frac{-1}{\varepsilon(E)}\right]\frac{E'dE}{E'^2 - E^2}$$

where P is the Cauchy principal part of the integral, avoiding the pole at E=E' [10,41]. Before reaching this point however, the experimental data must be treated in order to reduce the experimental spectrum down to the loss function. The experimental spectrum can be related to the single-scattering distribution (SSD) by accurately removing the zero loss peak and plural scattering. In this work, the Digital Micrograph function for Fourier-Log deconvolution was used to remove plural scattering while a fitted logarithm tail in the range 1.0 – 1.5 eV was used to remove the zero loss peak. Removing plural scattering may be unnecessary due to the monolayer MoS$_2$ being thin enough to discount plural scattering. The SSD can then be related to the loss function by [41]:

$$S(E) = A \cdot Im\left[\frac{-1}{\varepsilon(E)}\right] \cdot \ln\left[1 + \left(\frac{\beta}{\theta_E}\right)^2\right]$$

Where A is the normalisation factor, the middle term is the loss function and the latter term is the angular correction where $\beta$ is the collection semi-angle and $\theta_E$ is the characteristic scattering angle for a certain energy, E. The large convergence($\alpha$) and collection($\beta$) angles used in the experimental spectra acquired in this work may affect the accuracy of the angular corrections performed during the procedure. There are also momentum transfers outside of the dipole limit (q>0) being sampled which means a direct comparison with the dielectric function at the optic limit (q=0) is not accurate.

Attempts to determine the dielectric function from the experimental spectra have shown a large dependence on the normalisation factor used in the calculation. The built-in KKA function in Digital Micrograph uses a normalisation factor proportional to $\frac{1}{n^2}$ where n is the static refractive index. This method generally approximates a large value for n, e.g. 1000, for metals and high refractive index semiconductors. However this is not a reasonable approximation for monolayer MoS$_2$, and so a different normalisation factor is used based on the thickness of the medium:

$$A = \frac{2I_0 t}{\pi a_0 m_0 v^2}$$

where, $I_0$ is the integral of the zero-loss peak, t is the thickness, $a_0$ is the Bohr radius, and v is the speed of the incident electron. The effect of varying the thickness used in the normalisation factor shows a strong impact on the derived dielectric function. The Hyperspy KKA function gives the option to use thickness in normalising the SSD so this was used for a range of thickness measurements in Figure 3 to show the differences in derived dielectric functions. For thicknesses of (a) 0.3 nm and (b) 0.5 nm, the derived dielectric function seems to be unrealistic, where the real part shows behaviour similar to a metal with a negative value at lower energies, asymptotically approaching 0, whilst the imaginary part of the dielectric function seems to continue to resemble a semiconductor with a bandgap exhibiting the appropriate peaks for interband transitions. Then for thicknesses of (c) 0.7 nm and (d) 0.9 nm the derived dielectric functions start to look more typical of semiconductors. The real part of the dielectric function crosses zero near 2 and 3.5 eV which suggests possible plasmonic behaviour. After about 5 eV the real part levels out at 1 and the imaginary part drops to practically zero with no zero crossing for the $\pi$ and $\pi + \sigma$ plasmon peaks. When the thickness is increased further to (e) 1.1nm and (f) 1.3 nm the dielectric function starts to follow a trend where the real part approaches 1 and the imaginary part approaches. The real part of the dielectric function approaches 1 and no longer crosses zero at any energy, suggesting no collective excitation of a plasmon. The imaginary part of the dielectric function in (f) now closely resembles the EELS spectrum for monolayer in Figure 2. This resemblance is reported in previous studies of $MoS_2$ and is further observed in the simulation of dielectric functions from DFT [33,42].

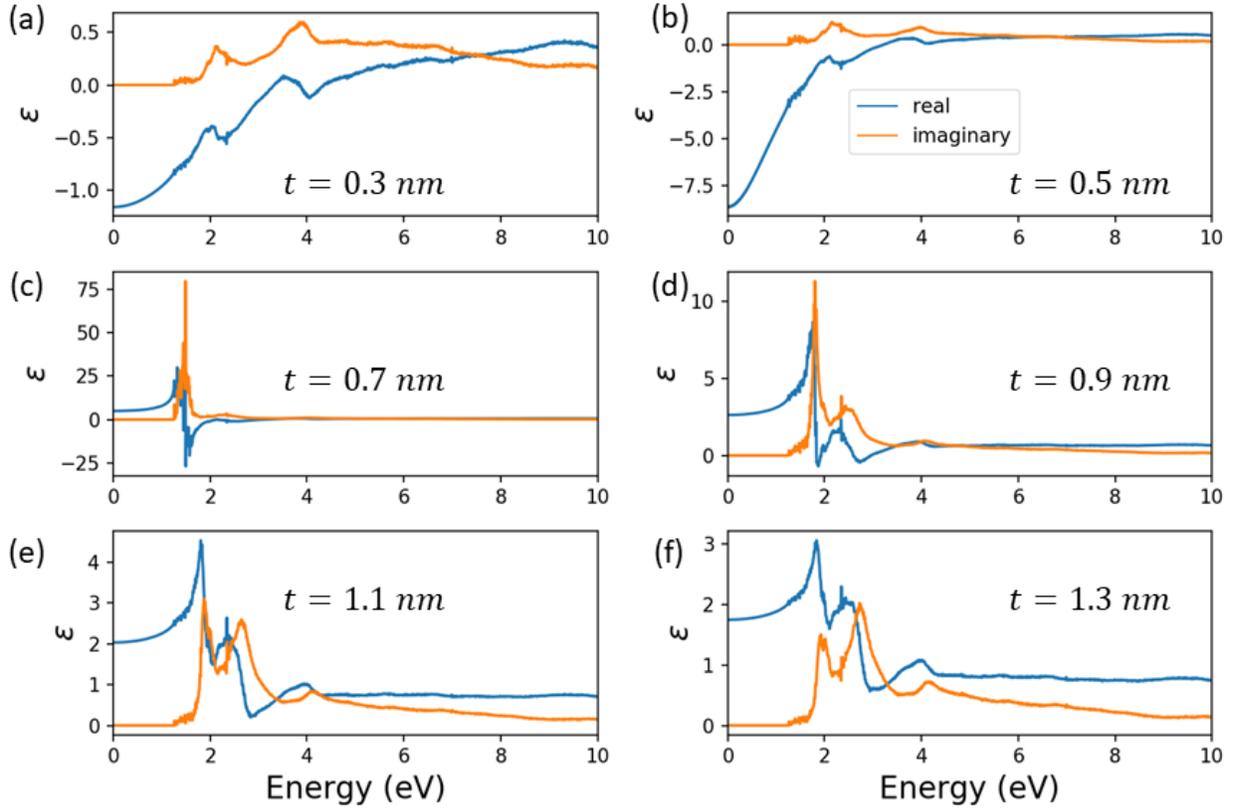

*Figure 3: Kramers-Krönig analysis of experimental EELS data with different thicknesses used in the normalisation.*

DFT simulations of EELS spectra rely on modelling the microscopic dielectric function, $\epsilon_{GG'}(\omega)$, for MoS$_2$ where G and G' denote reciprocal lattice vectors. This is then related to experimental EELS via the loss function, $-Im\left(\frac{1}{\epsilon_{00}(\omega)}\right)$, which contains local field effects. In order to model a monolayer of MoS2, the interlayer distance is increased by inserting a vacuum gap between the layers in the crystal. The size of the gap used is important to ensure that there is no coupling between the layers. Figures 1 and 2 show that the bulk plasmon modes completely disappear in the monolayer. The dielectric function is a bulk property of the material, therefore the loss function is also a bulk property. Surface losses in the experimental spectra can be determined by the differences between the bulk loss function and the experimental spectrum. . We observe that the real part of the dielectric function in Figure 4(a) of bulk MoS$_2$ (i.e. normal interlayer spacing) crosses zero at around 8 eV and 23 eV. These crossings are not present anymore in Figure 4(b) for the 119 Å interlayer distance, hence the criterion for a bulk plasmon is not met. This is also shown in Figure 5(a) where the loss function shows the decrease of the bulk plasmons as the layers are further removed from each other. The larger interlayer distance introduces a larger vacuum in between the layers. The system can be considered as a series of capacitors, and, by introducing this vacuum gap, the susceptibility, $\chi_{GG'}(\omega)$, decreases significantly. The dielectric function is related to the susceptibility by:

$$\epsilon(\omega) = 1 - v\chi(\omega)$$

Where, $v$ is the coulomb matrix. Since the susceptibility decreases with increasing interlayer distance, the dielectric function approaches 1. This behaviour is similar to that observed during the normalisation of experimental EELS spectra for KKA in Figure 3 where the dielectric function approached the vacuum conditions of $Re(\varepsilon) = 1$ and $Im(\varepsilon) = 0$.

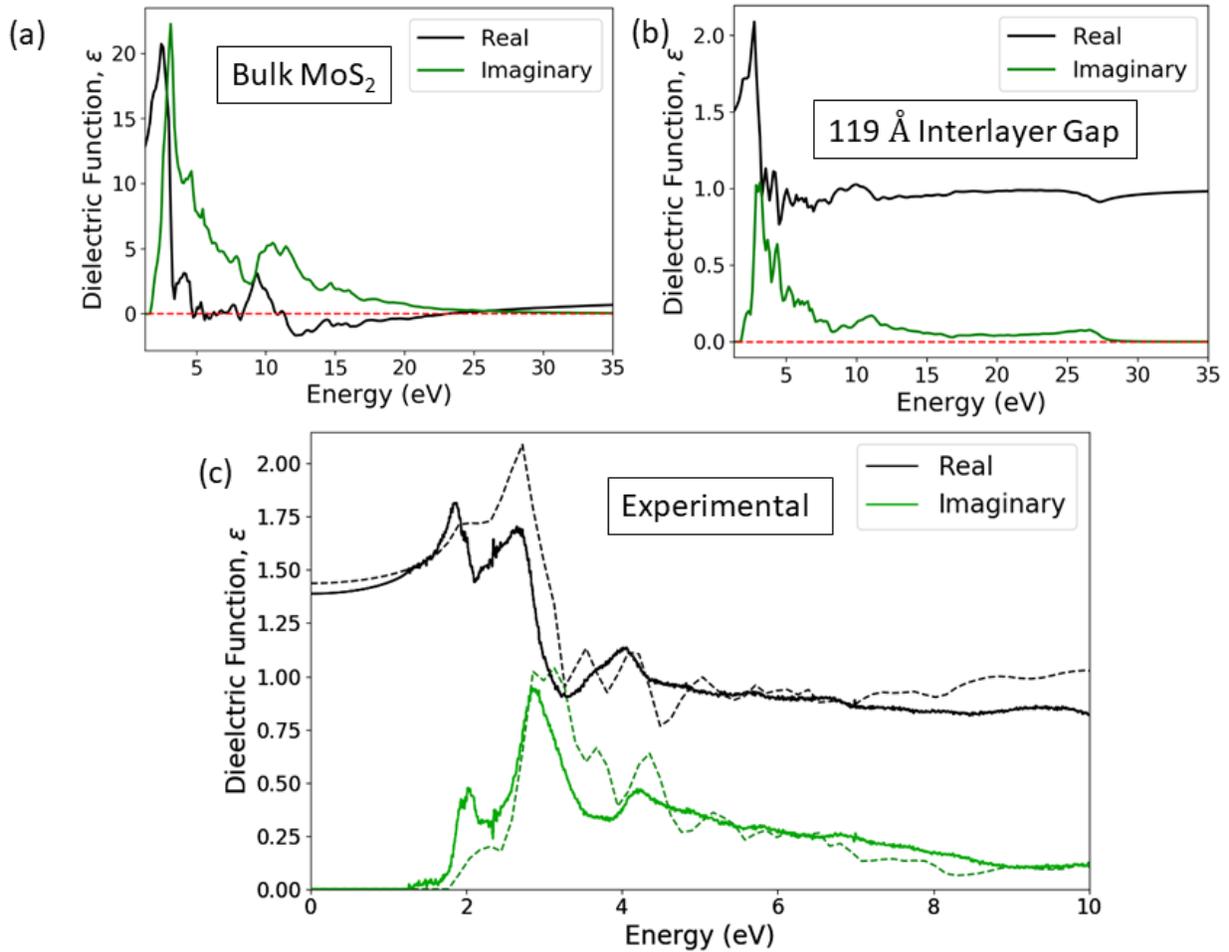

***Figure 4:*** *(a) Real and imaginary parts of the dielectric function of Bulk MoS$_2$ (b) Real and imaginary parts of dielectric function of MoS$_2$ when the distance between each layer in the crystal is increased to 119 Å to reduce the coupling between the layers. (c) Experimentally*

*determined dielectric function of MoS$_2$ extracted via Kramers-Krönig analysis of EELS spectra on monolayer MoS$_2$ using a thickness of 2 nm for normalisation. The simulated dielectric function with the 119 Å gap is overlaid as dashed lines.*

In Figure 5(b), the simulated energy loss spectrum resembles the experimentally acquired EELS spectrum quite well, except for a few details. The excitons in the experimental spectrum can be seen as sharp peaks in the 1.8 – 3.5 eV energy range while they are not clearly seen in the simulated spectrum. This is due to the usage of the random-phase-approximation (RPA) in which excitonic effects are neglected. The sharp peak at ~4 eV is an interband transition that can be seen clearly in both spectra and also corresponds to a similar peak in the imaginary part of the dielectric function as seen in Figures 4 & 5. The spectrum then up to ~8 eV can be characterized by a series of interband transitions, which again correspond to a series of peaks in the imaginary part of the dielectric function. The presence of the $\pi$ surface plasmon mode is obscured by these transitions and is difficult to pick out without further treatment through momentum resolved EELS. The lack of a zero crossing in the dielectric function could suggest that there is no $\pi$ surface plasmon in the monolayer. A similar claim has been made for graphene by Nelson et al. [43]. However, if there is a surface plasmon here then it could be heavily damped by interband transitions causing a lack of a clear peak in the spectra. Another interband transition at ~12 eV is also common to both experiment and theory. There is a clear difference between the experimental loss function and the simulated loss function in the range where we expect to see the $\pi + \sigma$ surface plasmon mode, indicated by the shaded region in Figure 5(b). This simulated loss function though is calculated for zero momentum transfer, $q = 0$. The converged STEM probe used in the experimental EELS allows contributions from non-zero momentum transfers so these must also be considered and further investigated. Finally a small peak at about 26 eV is only observed in the simulated spectra and is most likely the result of an interband transition as seen in the imaginary dielectric function in Figure 4(b).

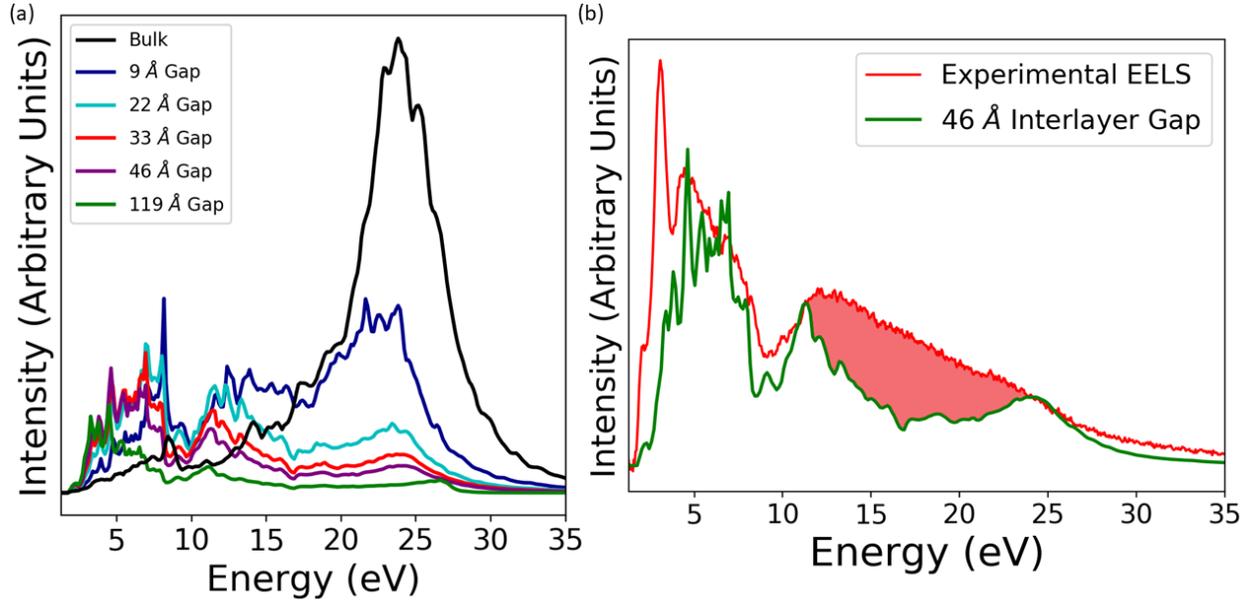

*Figure 5: (a) Energy loss function, $-Im\left(\frac{1}{\epsilon(\omega)}\right)$, from RPA for different interlayer distances. (b) Direct comparison of experimental EELS spectrum (red) and simulated spectrum with 46 Å interlayer gap. The experimental spectrum (red) is taken from the same monolayer area as shown in Figure 2 with the zero loss peak removed manually by the digital micrograph background subtraction tool.*

Momentum resolved measurements give more insight into the differences between the experimental and the theoretical loss function. Further simulations were done using an interlayer gap of 22.5 Å for a range of momentum transfers going from q=0 to 1.33 Å$^{-1}$ in the Γ → K direction. Figure 6 (a) and (b) show the imaginary dielectric function over a range of different momentum transfers, q. In the energy range where the presence of the $\pi + \sigma$ surface plasmon (12-20 eV) is expected, there are indirect interband transitions that blue shift in energy with increasing q. The transitions near q=0 are more intense but a weighted average of the simulated loss functions should resemble the experimental EELS spectrum. These indirect transitions contribute to the EELS signal in the shaded region of Figure 5(b) making it more difficult to isolate the $\pi + \sigma$ surface plasmon. Nerl *et al.* showed that the weight of the q=0 component, relative to the q>0 component, increases as a function of the accelerating voltage [33]. Finding the exact weighting factors could allow better isolation of the $\pi + \sigma$ surface plasmon mode. *Nerl et al.* also used the criteria of q=0 contributions to describe single-particle excitations and q>0 contributions to describe plasmons. It is possible that the peak at 14 eV in their spectra which they described as having plasmonic character could have been indirect transitions due to their large convergence and collection angles similar to this work. Mohn et al. [44] conducted similar momentum resolved measurements on MoS2 in the $\Gamma \to M$ direction.

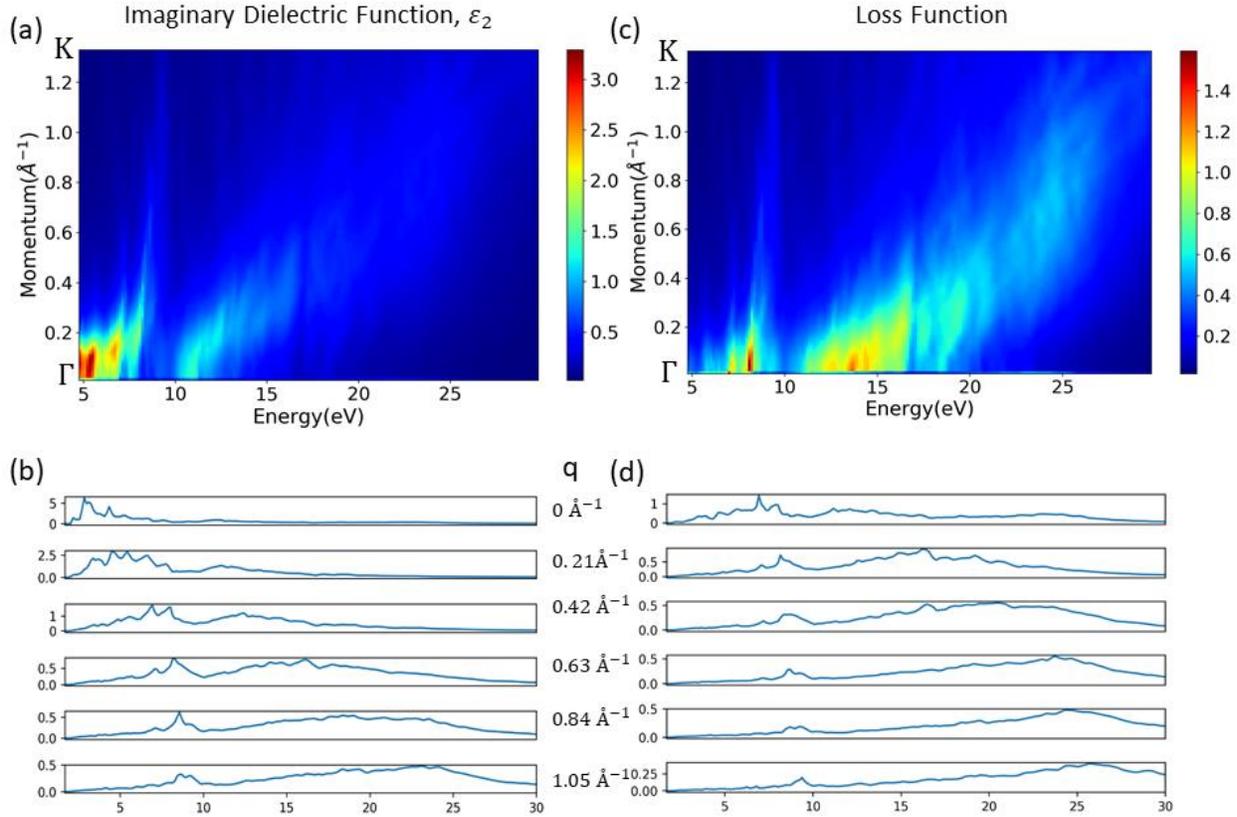

Figure 6: (a) Momentum resolved map of the imaginary part of the dielectric function, $\varepsilon_2$, in the $\Gamma \to K$ direction. (b) Plots of the peak shifts in $\varepsilon_2$ as a function of momentum transfer. The peaks in the $q \neq 0$ spectra show indirect interband transitions.

## Conclusions

The results shown in this work show clear differences in the behaviour of monolayer $MoS_2$ and graphene. The $\pi$ and $\pi + \sigma$ surface plasmons in monolayer $MoS_2$ are convoluted with single-particle excitations. A possible result of this is also increased damping of the plasmons in semiconducting $MoS_2$. Doping the monolayer with more carriers might increase the intensity of the surface plasmons as there are plasmons shown to exist at metallic edge states in $MoS_2$ nanostructures [32,45]. The Kramers-Krönig analysis of monolayer $MoS_2$ is shown to match with the DFT simulated dielectric functions for a large interlayer gap. However, the sensitivity of the normalisation step of KKA does imply that it would not be reliable without confirmation from a complimentary technique. DFT simulated dielectric functions for different momentum transfers show indirect interband transitions that contribute to the EELS spectra, which is the likely cause for higher intensities observed in the 12 to 26 eV regime in experimental spectra. This work provides the background necessary for understanding the response of pristine $MoS_2$, which is necessary for any attempts to functionalise the material.

## Acknowledgements

The authors would like to acknowledge the Irish Research Council Enterprise Scheme in conjunction with the Ernst Ruska-Centre, Forschungszentrum Jülich for funding the PhD work of Eoin Moynihan. The DFT simulations were supported through the "Integration of Molecular Components in Functional Macroscopic System" initiative of the VWStiftung. We gratefully acknowledge the computing time granted through JARA-HPC on the supercomputer JURECA at Forschungszentrum Jülich. Thanks to Dr. Jhih-Sian Tu for preparing the MoS2 sample used in this experiment.